\pdfoutput=1 
\documentclass[english,11pt,a4paper]{article}
\usepackage{jinstpub}

\usepackage{amsthm}
\usepackage{amsmath}
\usepackage{amssymb}
\usepackage{graphicx}
\usepackage[numbers]{natbib}
\usepackage{enumitem}

\title{B-spline parameterization of spatial response in a monolithic scintillation camera }

\author[a,1]{V.~Solovov\note{Corresponding author.},} 
\author[a,b]{A.~Morozov,} 
\author[a,b]{V.~Chepel,} 
\author[a]{V.~Domingos,} 
\author[a,b]{and R.~Martins}

\affiliation[a]{LIP-Coimbra,\\ Department of Physics, University of Coimbra, 3004-516 Coimbra, Portugal}
\affiliation[b]{Department of Physics, University of Coimbra,\\ 3004-516 Coimbra, Portugal}

\emailAdd{solovov@coimbra.lip.pt}

\abstract{
A framework for parameterization of the light response functions (LRFs) in a scintillation camera is presented. It is based on approximation of the measured or simulated photosensor response with weighted sums of uniform cubic B-splines or their tensor products. The LRFs represented in this way are smooth, computationally inexpensive to evaluate and require much less computer memory than non-parametric alternatives. The parameters are found in a straightforward way by the linear least squares method. Several techniques that allow to reduce the storage and processing power requirements were developed. A software library for fitting simulated and measured light response with spline functions was developed and integrated into an open source software package ANTS2 designed for simulation and data processing for Anger camera type detectors.}

\keywords{Scintillation detectors; Data processing methods; Simulation methods and programs}

\usepackage{babel}
\begin{document}
\maketitle

\section{Introduction}

The scintillation camera, first developed by H. Anger more then 50 years
ago \citep{anger_scintillation_1958}, remains an effective tool in
areas as diverse as medical imaging \citep{bushberg_essential_2011},
astrophysics \citep{Cook1985} and neutron detection \citep{ISI:000166992400002}.
Since the invention of the camera and until now the Center of Gravity
(CoG) algorithm remains the most widely used method for position reconstruction
of scintillation events. Its simplicity, and the fact that it can
be implemented using analog circuitry makes it the method of choice
despite several drawbacks, most notable being the distortion
of the reconstructed image, which is especially severe at the periphery of the
scintillator crystal. In modern commercial devices this distortion
is typically corrected by means of a look-up table \citep{Peterson2011}. The peripheral part of the crystal is not used, however, since image quality remains poor there even after the correction. Also, maintaining high
image quality requires frequent detector re-calibrations, considerably
increasing the running costs.

An alternative approach is to use Statistical Position Reconstruction (SPR) techniques \citep{Barrett2009}, \citep{hesterman_maximum-likelihood_2010} (e.g., based on maximum likelihood or least squares methods), that reconstruct an event by finding a set of parameters (typically, position and energy) for which the predicted detector response is in best agreement with the observed one. These techniques offer several advantages over the CoG method, such as a distortion-free image across the whole crystal area \citep{Vinke2014}, much better discrimination of noise and double events \citep{Hunter2009} and the possibility to reconstruct depth of interaction using flat sensor array \citep{Ling2007}. 

The SPR algorithms are computationally expensive, which for many years hampered their wider adoption. Nowadays, widespread usage of parallel processing hardware such as multi-core processors and graphics processing units (GPUs) made it possible to perform real-time position reconstruction using SPR algorithms on consumer-grade personal computers \citep{hesterman_maximum-likelihood_2010}. For example, a GPU-based SPR using an Nvidia GTX 770 GPU for a gamma camera with 37 photomultipliers was capable of processing up to $5\times10^6$ events per second \citep{Morozov2015}. It can be concluded that high computational power demand should not be regarded as a problem anymore.

In order to achieve accurate position and energy reconstruction with SPR techniques, a detailed characterization of the detector response (response of the light sensor array to the scintillation light as a function of the event position) must be performed beforehand. The traditional way of doing this is to scan the field of view of the detector with a pencil beam of gamma rays using a collimated gamma source. This is a time consuming process: for example, a scanning procedure reported in \citep{espana_digipet} covered a square grid of 32 x 32 nodes with an acquisition time of 160 seconds per node, or about 45 hours in total. Several techniques can be used to reduce the scan time. For example, a simultaneous scan with multiple sources using a coarser grid with cubic spline interpolation was successfully employed in \citep{miyaoka_cmice_cal}. 

It was also shown that spatial response of a scintillation camera can be reconstructed from two one-dimensional scans with linear sources in orthogonal directions \citep{van_dam_improved_nn} or, in some cases, from uniform gamma-ray irradiation data. For example, an iterative algorithm for reconstruction of the detector response was developed for the ZEPLIN III dark matter detector \citep{solovov_position_2012}. Later, it was successfully applied to the analysis of experimental data from several radiation detectors with optical readout similar to that used in Anger camera, such as LUX liquid xenon dark matter detector \citep{akerib_technical_2013}, a thermal neutron position-sensitive detector \citep{morozov_adaptive_2013} and a medical gamma camera \citep{Morozov2015}. A different calibration method, also using a non-collimated gamma source, was recently reported in \citep{espana_som}. In this method the mapping from the space of PMT response vectors to the space of event coordinates was created from a self-organizing map (SOM) constructed using the calibration data.

When developing software tools for statistical reconstruction, one is faced with the problem of storage of the detector response model. It can be stored either in a non-parametric form, for example, as a look-up table \citep{morozov_ants1}, or it can be parameterized as a set of light response functions (LRFs) that characterize the response of individual photosensors as functions of coordinates. The second approach has several advantages:
\begin{itemize}
\item substantially reduced storage requirements. For example, in \citep{morozov_adaptive_2013} the spatial response of every sensor was stored as a look-up table of ~1000 values. The parameterization approach described in this paper allows to reduce the number of parameters per sensor to 10 without any appreciable loss in the quality of the reconstructed image. This is especially important for processing on GPUs where the amount of fast memory is limited. 

\item the parameterization can also be used as a regularization technique that constrains the LRF to a required degree of smoothness, or ensures  symmetry or monotonicity of the the function. This regularization is particularly important for improving the stability of the iterative response reconstruction \citep{Morozov2015}.
\item the parameterized function typically has smooth coordinate dependence. This permits the use of more efficient minimization algorithms for SPR, for instance those which require the derivatives or Hessian functions of the LRFs \citep{Joung2002}.
\end{itemize}

Our group is developing a comprehensive set of software tools for scintillation
cameras, including simulation, detector response reconstruction, data processing, event reconstruction and analysis, distributed as an open source software package ANTS2 \citep{morozov_ants2}. This required the development of a technique to store the detector response model that (i) can be used with  detectors of arbitrary geometry, (ii) permits fast evaluation, (iii) has moderate storage requirements and (iv) can be used for unsupervised fitting of the simulated and experimental data. Several options were considered: analytic representation, Taylor series, Fourier series and splines. The  analytic representation, where the LRF is expressed as a custom function of coordinates with a few adjustable parameters, does not satisfy requirement (i) as the adequate function is always geometry-specific. To achieve satisfactory results with the Taylor series, the degree of the polynomial must be fine tuned on a case-by-case basis to avoid both under and over-fitting. This consideration made it incompatible with requirement (iv). The evaluation of the Fourier series involves computation of trigonometric functions and for this reason was considered too slow. Spline parameterization, widely used in engineering applications for interpolation and data smoothing, was found to satisfy all the criteria listed above: it is flexible enough to handle arbitrary detector configuration while preserving all the advantages of the parametric approach.

The considerations above led to the development of a software library for fitting the photosensor response using splines in one and two dimensions, along with the design of several techniques for reducing the total number of parameters by taking into account the spatial symmetry of the photosensor response. This paper features the implementation details, as well as a few application examples.

\section{Splines}

Splines in general are a class of piecewise polynomial functions with continuous derivatives up to a certain degree. Splines made of higher degree polynomials are smoother but more costly to evaluate. Following the approach used in \citep{miyaoka_cmice_cal}, \citep{solovov_position_2012} and \citep{Joung2000}, we have decided to use cubic splines for LRF parameterization as it ensures a smooth first derivative (important for convergence of several minimization algorithms) while still not being too computationally expensive. Besides piecewise polynomial functions, splines can also be presented as a weighted sum of basis functions called B-splines. While both these approaches are mathematically equivalent \citep{Unser1999}, the B-spline representation requires less storage (by almost a factor of 4 in the case of cubic splines), as will be discussed in Section \ref{sub:Storage and evaluation}.

\subsection{Uniform cubic B-splines}

\begin{figure}[h]
\begin{center}
\includegraphics[width=7cm]{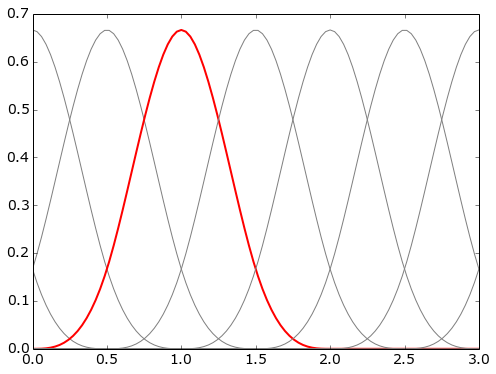}
\includegraphics[width=7cm]{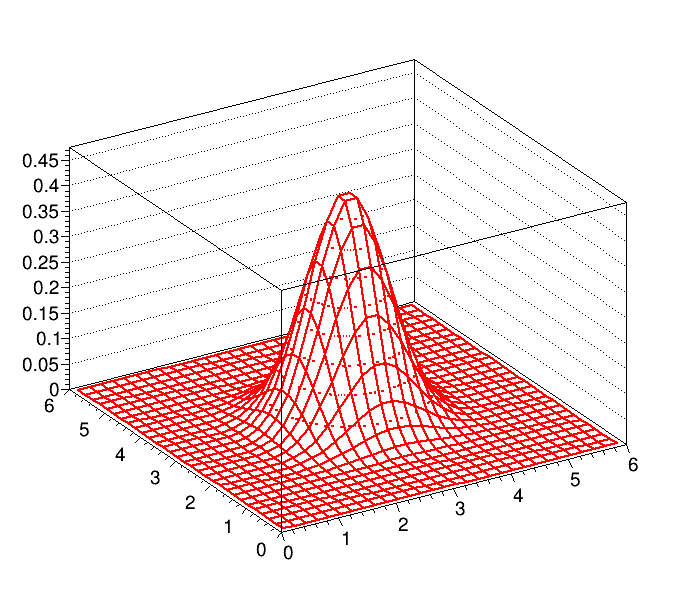}
\end{center}

\caption{\label{fig:UCBS}An example of the full basis formed by uniform cubic B-splines on the segment $[0,3]$ (left). Two dimensional tensor product of two UCBS (right). The labels on the horizontal axes coincide with the knot positions.}

\end{figure}

B-splines (basis splines) are a set of functions that form a
basis for the spline functions of a given degree over a certain interval
$[x_{0},x_{1}]$, i.e. any spline $S(x)$ can be represented as a linear combination
of B-splines:
\begin{equation}
S(x)=\sum_{k}w_{k}B_{k}(x)\label{eq:Bslpine_sum}\quad,
\end{equation}

where $B_{k}(x)$ are the B-splines and $w_{k}$ are the spline coefficients.
Each B-spline of n-th degree consists of n+1 polynomial segments separated
by control points called knots. The shape of a B-spline is defined
by its degree and the placement of the knots. In the case of uniform B-splines the knots are spaced with equal intervals and all $B_{k}(x)$ are shifted copies of each other:

\begin{equation}
B_{k+1}(x)=B_{k}(x-d)
\end{equation}

or, equivalently:
\begin{equation}
B_{k}(x)=B_{1}(x-(k-1)d),\label{eq:B_k_shift}
\end{equation}

where $d$ is the interval between the knots. In the case of uniform cubic B-splines (UCBS), a B-spline consists of four cubic segments arranged to ensure continuity of the second derivative as shown by the thick red curve in the left plot of Figure \ref{fig:UCBS}. The same figure also shows an example of the full basis made of B-splines for a segment (in this case $[0,3]$). Please refer to \citep{Unser1999} and \citep{Brunet2010} for a detailed description of B-splines and their properties.

 While non-uniform splines offer greater flexibility, they require additional storage for the knot positions. They are more expensive to evaluate as well, because before computing the polynomial, the corresponding segment must be found using a binary search.  For these reasons, it was decided to base the LRF parameterization on UCBS. 

\subsection{Tensor product splines}

B-spine representation can be expanded to two or more dimensions using tensor
products. For example, in the 2D case one can write, similarly to \ref{eq:Bslpine_sum}: 
\begin{equation}
S(x,y)=\sum_{k}\sum_{l}w_{kl}B_{k}(x)B_{l}(y)\label{eq:tp_spline}
\end{equation}
Here $B_{k}(x)B_{l}(y)$ is a product of one-dimensional B-splines in the \(x\) and \(y\) directions. A single two-dimensional tensor product of UCBS is shown on Figure \ref{fig:UCBS} (right). It consists of 16 bi-cubic segments and
retains attractive features of one-dimensional B-splines, such as
smoothness of the first derivatives and compact support. 

\section{LRF parameterization}

The LRF characterizes the spatial dependence of the response of a given photo sensor in a scintillation camera. It can be defined everywhere inside the scintillator volume and is generally a function of three coordinates. However, in many practical applications (e.g. most gamma cameras and SPECT detectors \citep{Peterson2011}, as well as dual phase noble fluid time projection chambers with optical readout \citep{aprile_2013}, \citep{darkside_2013}, \citep{apd_el_tpc}), the scintillator is thin enough for the dependence of the LRFs on the depth of interaction \(z\) to be negligible, thus making it possible to consider them as only being dependent on \textit{x} and \textit{y}. We will assume that this is the case for the rest of this discussion. 

\begin{figure}[h]
\begin{center}
\includegraphics[width=7cm]{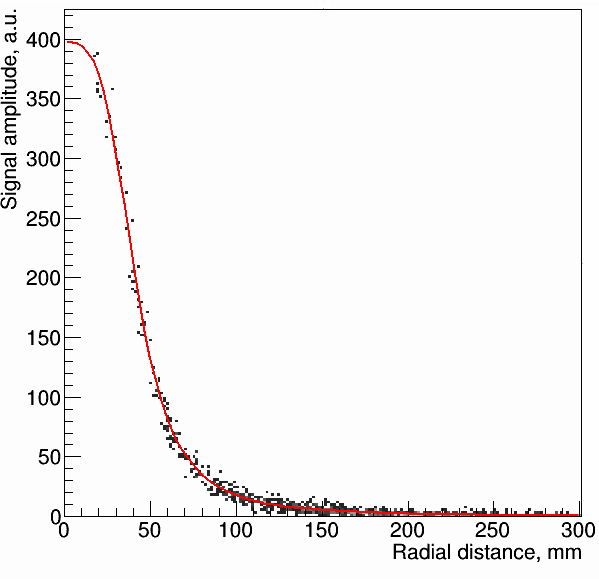} 
\end{center}

\caption{\label{fig:LRF_fit}Example of parameterization of axially symmetric response of a photosensor (PMT in a medical gamma camera). The points show signal amplitude of one of the PMTs plotted against the distance from the simulated event positions to the axis of the PMT. The line is the fit of the points with a cubic spline.}
\end{figure}

Typically, an event in a scintillation camera (a gamma ray interacting with the scintillator) is recorded as a vector of signal amplitudes registered by the photosensor array. In the following discussion, only events resulting in scintillation from a single point are considered. For an event producing in total \(N\) isotropically emitted photons, due to the statistical nature of photon detection, each element of this vector is a random number distributed with an expectation value of
\begin{equation}
\langle A_i \rangle = NC_i\eta_i(x,y)\label{eq:A_i_expectation},
\end{equation}
where \(A_i\) is the signal amplitude, \(\eta_i\) is the LRF and \(C_i\) is the scaling factor (gain), all for the i-th photosensor. The gains \(C_i\) become important when using the same LRF for several photosensors as described in section \ref{sub:Array symmetry and sensor groups}. To find the LRF, one needs a set of data points, where each data point contains the photosensor signals, the event coordinates and the number of emitted photons. The function \ref{eq:A_i_expectation} is fitted to this dataset to obtain the LRF parameters that most closely describe the available data as shown in Figure \ref{fig:LRF_fit}. 

Well implemented parameterization should satisfy the following criteria: 
\begin{enumerate} [label=(\alph*)]
\item faithful representation of the LRF. The deviations from the true function will create bias in the reconstructed position which in turn will lead to distortions in the reconstructed image. The tolerance to such deviations depends on the application but as a rule of thumb one can assume that the deviations can be considered acceptable if they are significantly smaller than statistical fluctuations of the photosensor signals;
\item optimal number of parameters, i.e. the minimum number of parameters that is required to satisfy property (a). Minimizing the parameter count, besides reducing the storage requirements, also leads to less data points required for the fit, less susceptibility to overfitting as well as a faster fit;
\item stable unsupervised fitting.  This is important for implementation in a software package not focused on a particular detector, especially if the package is supposed to perform parameterization in unsupervised mode.
\item fast evaluation. The LRF evaluation is the most often called routine (typically ranging from hundreds to thousands of times) for each reconstructed event, so its efficiency directly influences the time required for reconstruction. 
\end{enumerate}

\subsection{Use of axial symmetry\label{sub:Use of axial symmetry}}

Generally, an LRF must be fitted by a function of two variables (Equation \ref{eq:tp_spline}) to correctly represent its shape. However, in many practical cases such as the ZEPLIN III detector \citep{solovov_position_2012} or clinical gamma cameras \citep{Morozov2015}, the LRFs can be considered axially symmetric for the purpose of event reconstruction. If one assumes the response of a photosensor to be axially symmetric, this makes it a function of only one variable \textit{r}, which is the distance from the photosensor axis. In this case the LRF can be represented by a spline (Equation \ref{eq:Bslpine_sum}) as a function of \textit{r}:
\begin{equation}
\eta(x, y)=\sum_{k}w_{k}B_{k}(r)=\sum_{k}w_{k}B_{k}(\sqrt{(x-X_0)^2+(y-Y_0)^2})\label{eq:LRF_axial},
\end{equation}
where $\eta(x, y)$ is the LRF and $X_0$ and $Y_0$ are the coordinates of the photosensor axis. 

The use of axially symmetric LRFs, when applicable, significantly reduces the required number of parameters and speeds up LRF evaluation. Also, in the case of iterative reconstruction of the LRFs from the flood data, the axial symmetry provides sufficient regularization to ensure convergence even at low photon statistics \citep{Morozov2015}. 

The most straightforward case for use of axially symmetric LRFs  is a photomultiplier tube (PMT) with a circular photocathode. It can also be applied to hexagonal and even square photosensors, provided that the lightguide thickness is comparable with the sensor size, as demonstrated in section \ref{section:Examples}.

\begin{figure}[h]
\begin{center}
\includegraphics[height=5cm]{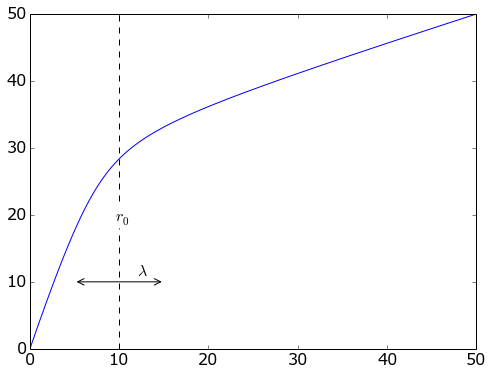} 
\includegraphics[height=5cm]{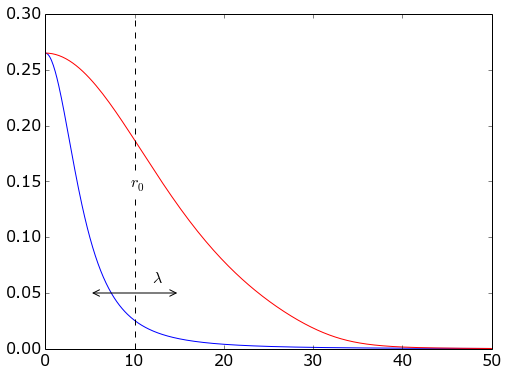}
\end{center}

\caption{\label{fig:plot_compression}Left: nonlinear scale transformation from \( r\) to \(\rho\); right: the same LRF  shown as a function of \( r\) (blue curve) and \(\rho\) (red curve).}
\end{figure}

\subsection{Compressed axial LRFs\label{sub:Compressed axial LRFs}}

Typically, an LRF $\eta(x, y)$ exhibits a much higher variation rate close to the corresponding photosensor center than it does at large distances. This results in a conflict when optimizing the number of knots for a uniform spline: on the one hand, the inter-knot distance must be short enough to faithfully represent the region of rapid function change, while on the other hand following this criterion leads to too many knots covering the slowly changing region. Consequently, for a large photosensor array most of the spline storage is used with little efficiency. Moreover, an excessive number of knots may cause LRF overfitting and the subsequent appearance of artifacts in the reconstructed images. An obvious solution would be to apply non-uniform splines, however the associated storage and computational overheads were considered prohibitive. 

An alternative solution was implemented for axially-symmetric LRFs \(\eta(r)\) in our parameterization library. A new variable \(\rho(r)\) was introduced:
\begin{equation}
\rho=a\Big({{\kappa+1}\over{\kappa-1}}(r-r_0)-\sqrt{(r-r_0)^2+\lambda^2}+b\Big)\label{eq:transform}
\end{equation}
It depends on \(r\) and varies faster for smaller \(r\) (closer to the photosensor) and slower for larger \(r\). The ratio between the variation rates at the origin and at infinity is defined by the constant \(\kappa\), while \(r_0\) and \(\lambda\) define the center point and smoothness of transition between the two regions (Figure \ref{fig:plot_compression}, left). Constants \(a\) and \(b\) are chosen to ensure that \(\rho\) covers the same domain as \(r\). An LRF \(\eta(\rho)\), as a function of this new variable, is stretched close to the origin and compressed at large distances, as shown in Figure \ref{fig:plot_compression}, right. After such a transformation (referred later as \emph{compression}, as it "compresses" the slowly changing LRF tail), \(\eta(\rho)\) requires less knots (and, therefore, less storage) than the original \(\eta(r)\) to fit experimental or simulated data with the same precision (see the second example in Section \ref{section:Examples}). The particular function \ref{eq:transform} was chosen because the user-defined parameters \(\kappa\), \(r_0\) and \(\lambda\) are easy to choose empirically from the observed shape of \(\eta(r)\), while \(a\) and \(b\) can be computed automatically.

\subsection{General 2D LRFs \label{sub:General 2D LRFs}}

There are detector configurations where the light response is not symmetric for at least some of the photosensors. One example of this is a sensor close to the side boundary of the crystal of a compact gamma camera, as shown in Figure \ref{fig:camera_model}. For such cases, the LRF must be fitted by a function of two variables (see Eq. \ref{eq:tp_spline}) to correctly represent its shape. The number of knots in both directions should be carefully optimized as the parameter count grows proportionally to the product of the number of knots, in turn resulting in a rapid increase in the storage requirements and in the time necessary for fitting.

\subsection{Array symmetry and sensor groups\label{sub:Array symmetry and sensor groups}}

It is possible to take advantage of the symmetry of the photosensor array geometry and use a common LRF for several sensors. This is done by arranging all sensors into groups, so that the same LRF can be used for all sensors in one group by defining the proper coordinate transforms, consisting of a combination of rotation, translation and  reflection. This approach not only reduces the storage requirements but also effectively multiplies the number of events available for LRF fitting by the number of sensors in the group. To account for the differences in the individual channel sensitivity, the gain parameter \(C_i\) is assigned to each sensor in the group. To calculate the expected response of a particular photosensor, the common group LRF is scaled with this gain parameter according to equation \ref{eq:A_i_expectation}.

In the extreme case of such grouping, a single axial LRF for all photosensors can be used. Despite the simplicity, it works well for geometries where the fraction of reflected light is negligible \citep{solovov_position_2012}. It was also found useful for the iterative LRF reconstruction in a medical gamma camera \citep{Morozov2015}.There it was used for the first few iterations before switching to iterations with individual LRFs to refine the result. 

\begin{figure}[h]
\includegraphics[width=4.5cm]{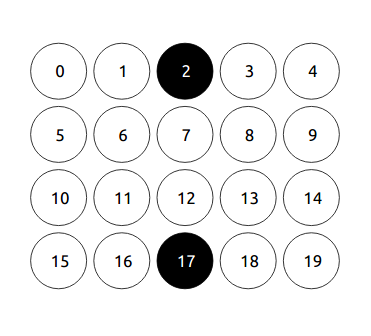}(a) 
\includegraphics[width=4.5cm]{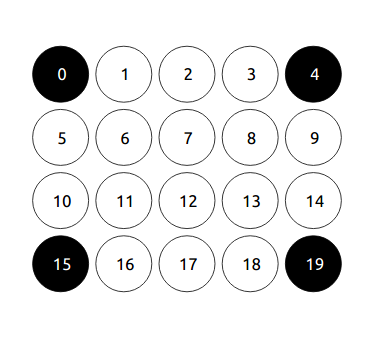}(b)
\includegraphics[width=4.5cm]{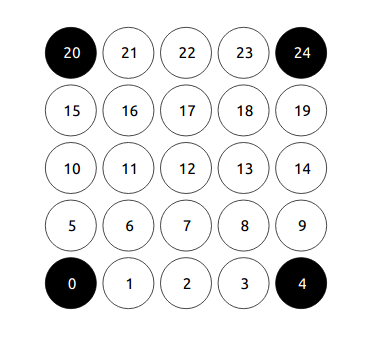}(c)
\includegraphics[width=4.5cm]{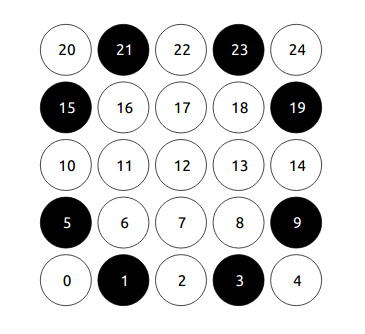}(d) 
\includegraphics[width=4.5cm]{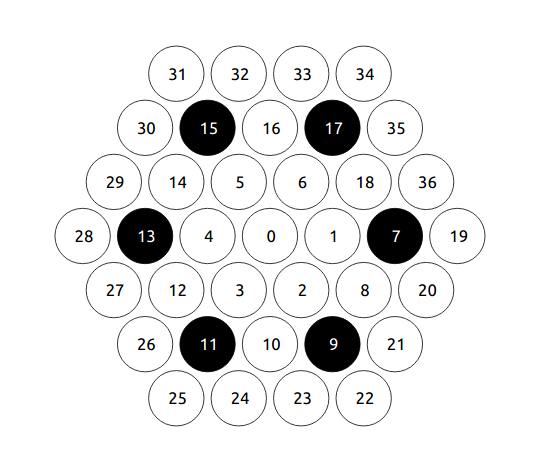}(e)
\includegraphics[width=4.5cm]{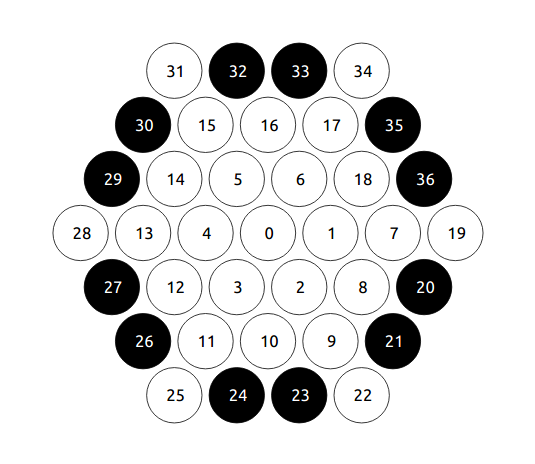}(f)
\caption{\label{fig:3_grouping}Examples of groups based on rotational and reflective symmetry in rectangular (a, b), square (c, d) and hexagonal (e, f) arrays of photosensors.}
\end{figure}

More complex grouping can be adopted by using rotational or reflection symmetry of the photosensor array. This allows the sensors to be arranged in groups of 2 and 4 for a rectangular array, groups of 4 and 8 for a square array and groups of 6 and 12 for a hexagonal one as shown in Figure \ref{fig:3_grouping}.

\section {Implementation details}

\subsection{Fitting of the data\label{sub:Fitting of the data}}

The spline coefficients $w_{k}$ are computed using a linear least squares fit to the experimental or simulated dataset made of records containing photosensor signals, the corresponding event coordinates, and the number of emitted photons (estimated in the case of experimental data). To make the fitting more efficient, the data points are binned: while it is technically possible to fit all the data points directly, the time required for the fit becomes impractical for datasets of more than a few thousand points. The binning is done by radius from  the photosensor for axial LRFs or by the \((x,y)\) position otherwise. Our library uses 4 bins per spline interval, which proved to be adequate for the detector geometries we have used so far. The \(w_k\) are then found by solving the over-determined system of linear equations \(\sum_{k}B_{kj}w_k=z_j\), where \(B_{kj}\) is the value of \(k\)-th B-spline in the middle of \(j\)-th bin and \(z_j\) is the photosensor response averaged over all points that fall into the \( j\)-th bin. In the case of axially-symmetric LRF, extra equations are added to force the derivative at the origin to be zero. Depending on the user's choice, the solution can be found by either QR-decomposition or singular value decomposition (SVD). However, even when a more robust SVD is used, the fitting is not particularly tolerant to the presence of empty bins, so care should be taken to ensure that the whole scintillator area is represented by the data points. 

When the photosensors are grouped as described in \ref{sub:Array symmetry and sensor groups}, a single LRF is fitted for all photosensors in a group. In this case each of the data points is "cloned" using the same coordinate transform that is applied to calculate the LRF of the group members, effectively multiplying the number of points available for the fit by the number of photosensors in the group.

\subsection{Storage and evaluation\label{sub:Storage and evaluation}}
The fitted spline can be stored in two formats, depending on the intended usage. Storing it as a set of coefficients  \(w_k\) is most efficient in terms of computer memory usage, requiring \(n+3\) variables per LRF, where \(n\) is the number of intervals. However, this format is more computationally expensive, since this case requires the LRF to be found as a sum of 4 cubic polynomials. An alternative approach, based on the fact that the spline itself is a piecewise polynomial, is to  pre-compute and store polynomial coefficients for each interval. In this case only one cubic polynomial has to be computed but \(4n\) values must be stored per LRF. These differences are even more pronounced for the tensor product splines. In the ANTS2 package, the first method is used for reconstruction on GPU, where the amount of fast memory is limited, while the second one is applied for reconstruction on CPU, where storage is not a limiting factor.

\section{Examples\label{section:Examples}}

\begin{figure}[h]
\begin{center}
\includegraphics[width=6cm]{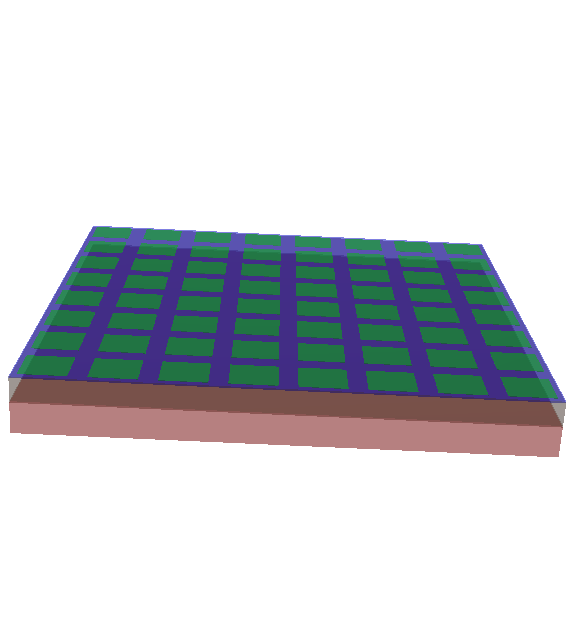}
\includegraphics[width=7cm]{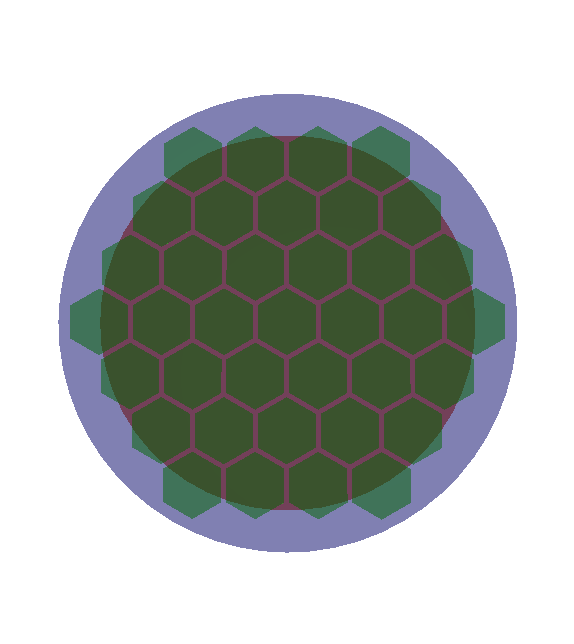}
\end{center}
\caption{\label{fig:camera_model} The example gamma camera models: a compact camera with a LYSO crystal and SiPM readout (left), and a gamma camera of standard configuration with a NaI(Tl) crystal and PMT readout (right).}

\end{figure}

In this section we demonstrate the viability of the technique described above on two simulated examples. The first of these examples illustrates event reconstruction in the compact scintillation camera shown in Figure \ref{fig:camera_model} (left). It consists of a 33.2$\times$33.2$\times$2.0 mm$^{3}$ LYSO scintillator crystal (pink) coupled to an 8$\times$8 array of 3$\times$3 mm$^{2}$ silicon photomultipliers (SiPMs) (green) through a 1.5 mm thick light guide (blue). The SiPM array comprises four identical elements modeled after the SensL ArrayC-30035-16P-PCB array \citep{SensL_Array}, including dimensions, photon detection efficiency, noise and micro-cell count. The simulations and reconstructions were performed using the ANTS2 package. The spline coefficients of a chosen parameterization scheme were calculated using the dataset emulating uniform irradiation by mono-energetic gamma rays, consisting of $5\times 10^{5}$ randomly placed scintillation events, uniformly distributed in the scintillator volume.
Each scintillation event was simulated as an isotropic emission of 4000 optical photons (this roughly corresponds to the number of scintillation photons produced by a 140 keV gamma ray in LYSO) from a given point inside the scintillator. Each photon was traced, taking into account the optical properties of the materials and their interfaces, until it was either detected by one of the SiPMs or lost. Each record in the dataset consisted of the event position and the vector of number of photons detected by each SiPM.
The obtained LRFs, in turn, were used to reconstruct the event positions via the method of maximum likelihood. To assess the performance of the reconstruction, two methods were used. In the first one, the scintillation events were simulated on a square grid with 2mm pitch and the density maps of reconstructed positions were visually inspected. In the second, in order to quantify the non-linearity of the reconstructed images, the average deviation of the reconstructed event position from the true one along the $x$ axis was calculated for another dataset, which was simulated with uniform irradiation. 

\begin{figure}[h]
\begin{center}
\includegraphics[width=7cm]{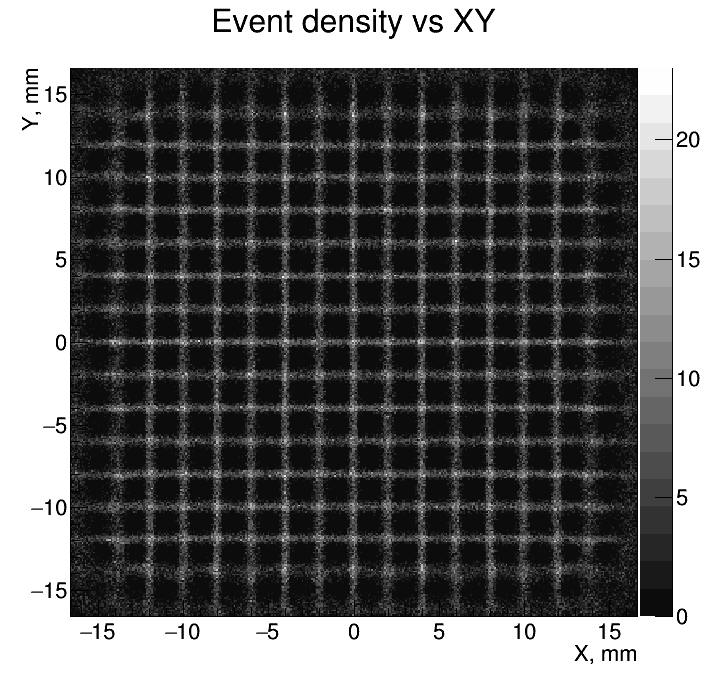}
\includegraphics[width=7cm]{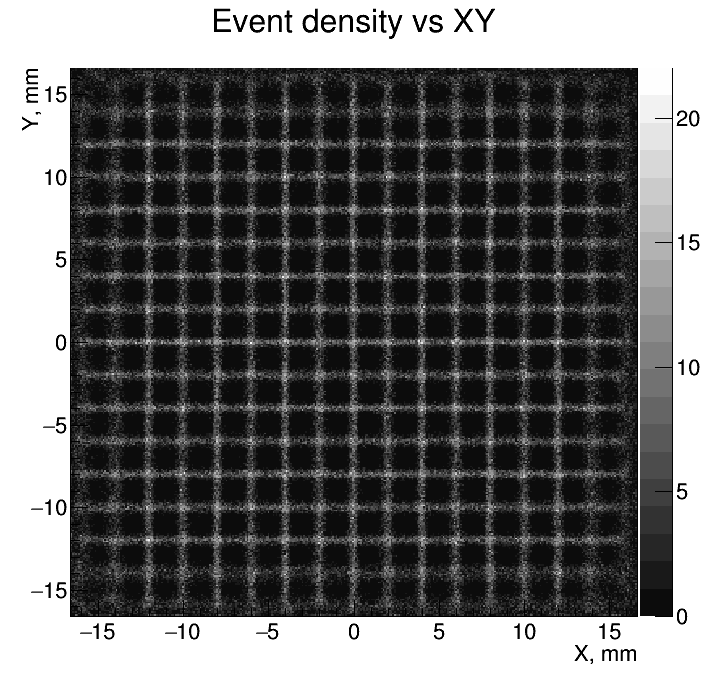}
\end{center}
\caption{\label{fig:square_camera_rec}Density maps of reconstructed positions of the events simulated on a square grid with 2 mm pitch for axially symmetric (left) and 2D (right) LRFs.}
\end{figure}

\begin{figure}[h]
\begin{center}
\includegraphics[width=7cm]{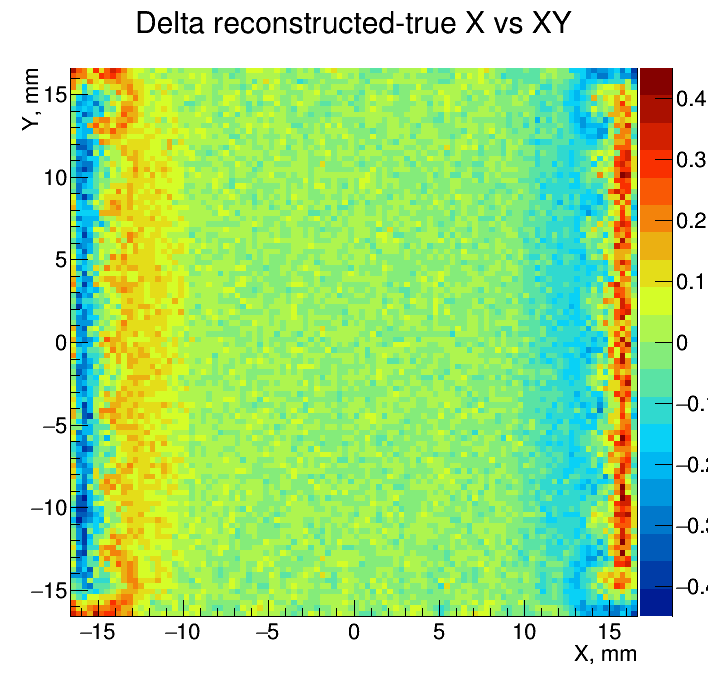}
\includegraphics[width=7cm]{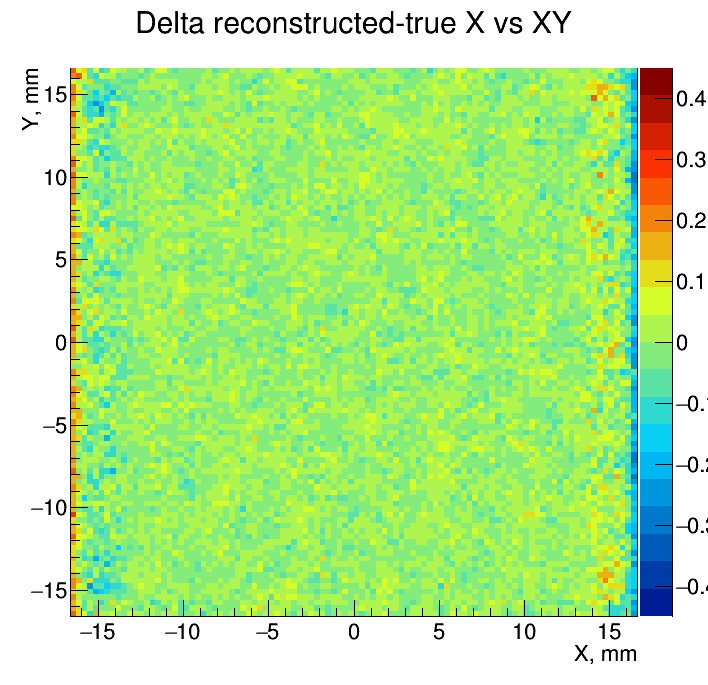}
\end{center}
\caption{\label{fig:square_camera_dx}Maps of the average deviation (in mm) of the reconstructed position from the true one along the $x$ axis for axially symmetric (left) and 2D (right) LRFs.}
\end{figure}

The density maps of the reconstructed positions for the events simulated on a 2$\times$2~mm grid  are shown in Figure \ref{fig:square_camera_rec}. The left image was obtained using axially symmetric functions to parametrize the LRFs of the SiPMs in the detector.  While this approach may seem counter-intuitive due to the square shape of the SiPMs, it works very well for the internal part of the camera. The reconstruction only fails near the border of the crystal, where the reflections from the side walls break the axial symmetry of the light response. This is illustrated by the lines at $\pm$14~mm that are noticeably pulled towards the center and the lines at $\pm$16~mm that practically disappeared. The image on the right was obtained using a general 2D LRF parameterization with 25 spline intervals in each direction. As one can see, there is a notable improvement in the border region while the internal part is essentially identical to that in the left image made with axially symmetric LRFs. The maps of the average deviation shown in Figure \ref{fig:square_camera_dx} confirm this observation: for axially symmetric LRFs, the absolute reconstruction bias is below 0.1 mm in the region between $-10$ and 10~mm, and below 0.2~mm for the events occurring between $-13$ and 13~mm. Outside the central 26~mm$\times$26~mm region, a bias of up to $\pm$0.45~mm can be observed. However, for 2D LRFs the bias is below 0.1~mm practically for the entire crystal area, excluding just about 1mm at the edge. Note that due to the symmetry of the gamma camera, this analysis holds for the $y$ direction as well.      

The second example demonstrates how the quality of the reconstructed image in a medical gamma camera can depend on the choice of LRF parameters, namely the number of spline intervals and the use of compression (see Section \ref{sub:Compressed axial LRFs}). It is based on the model of the camera used to investigate iterative LRF reconstruction \citep{Morozov2015}. This camera consists of a \o 470 mm NaI(Tl) scintillator crystal and a \o 570 mm glass lightguide, both 12.5 mm thick. The readout is done by an array of 37 hexagonal PMTs coupled to the lightguide (see Figure \ref{fig:camera_model} (right)). In this case, the simulation included the irradiation of the camera with a collimated beam of 140 keV $\gamma$-rays, followed by the propagation of scintillation photons. The flood data, obtained by uniform camera irradiation, was used to calculate the LRFs. To obtain the test data (used for the assessment of the reconstruction quality) a mask made of lead bars (10 mm wide and 3 mm thick) was placed in the beam path in front of the camera. Further details on the geometry and simulation parameters of the gamma camera can be found in \citep{Morozov2015}. 

\begin{figure}[h]
\begin{center}
\includegraphics[width=4.5cm]{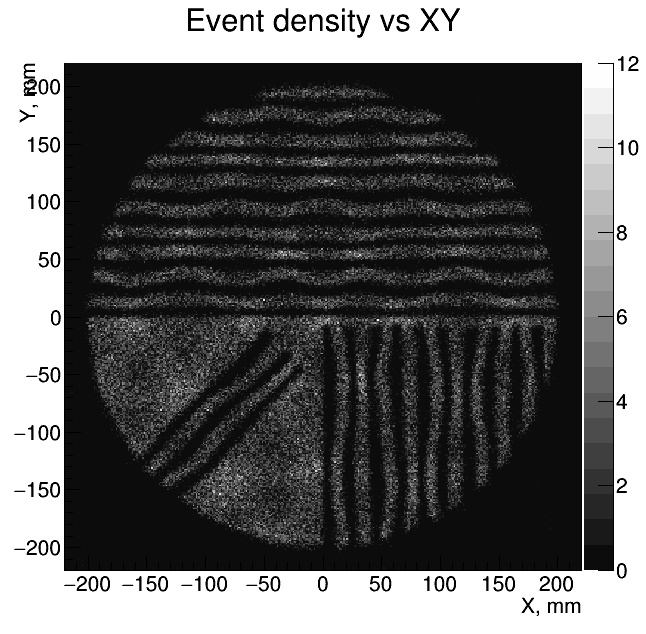} 
\includegraphics[width=4.5cm]{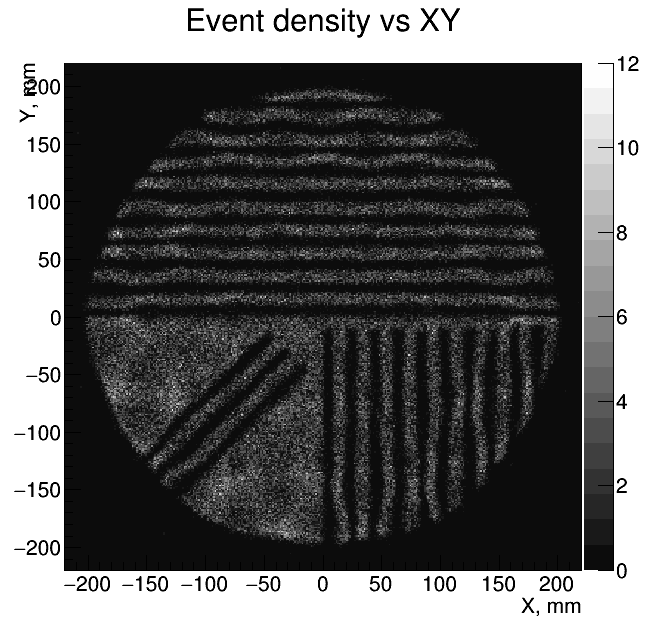}
\includegraphics[width=4.5cm]{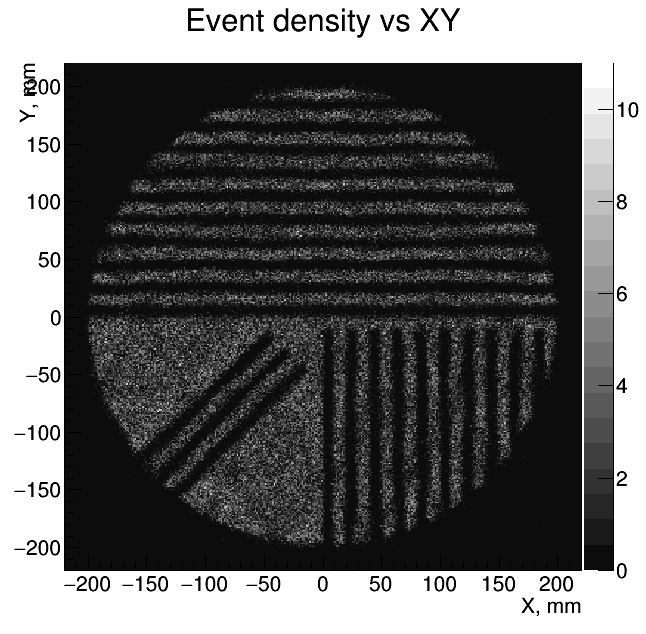}
\end{center}
\begin{center}
\includegraphics[width=4.5cm]{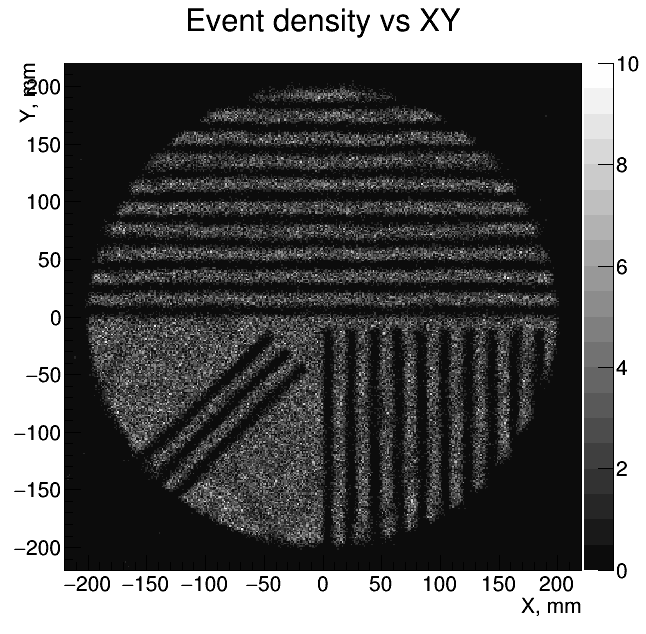} 
\includegraphics[width=4.5cm]{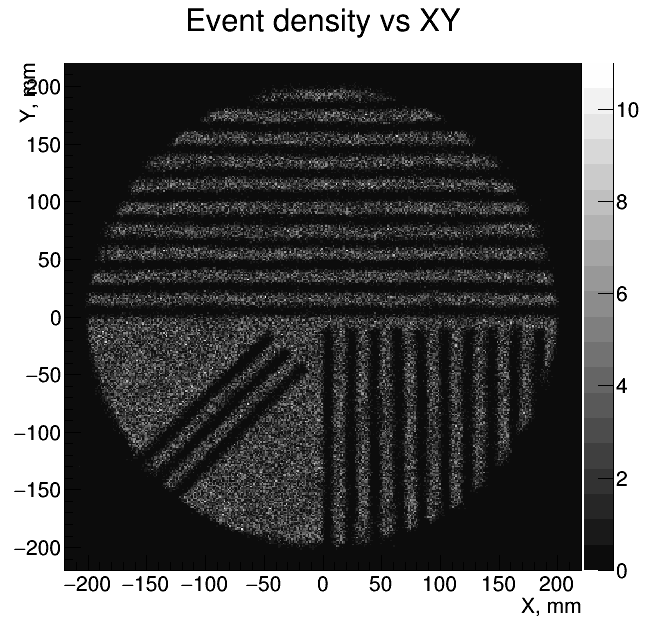}
\includegraphics[width=4.5cm]{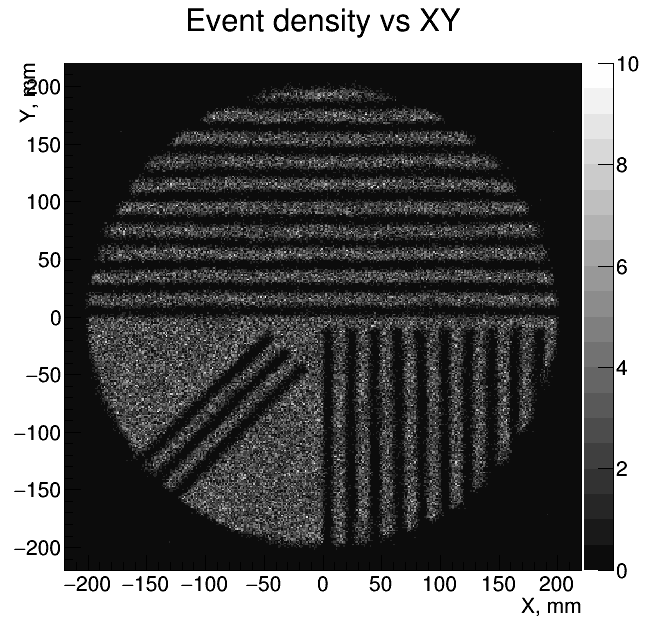}
\end{center}
\caption{\label{fig:masks}Mask image reconstructed using axial LRF parameterization with 8 (left), 12 (middle) and 16 (right) intervals per spline. Top row: no compression, bottom row: compression with $\kappa=5$, $r_0=150$ mm and $\lambda=50$ mm.}
\end{figure}

The images of the mask reconstructed using axial LRF parameterization are shown in Figure \ref{fig:masks} for (from left to right) 8, 12 and 16 intervals per spline. The top row of images was obtained using uniform splines without compression while the images in the bottom row employed the radius transformation \ref{eq:transform}. As one can see, the use of compression results in a considerably less distorted image for an equal number of intervals. In fact, compressed 8-interval and not compressed 16-interval representations yield reconstructed images with a similar degree of distortion.

\begin{figure}[h]
\begin{center}
\includegraphics[width=4.5cm]{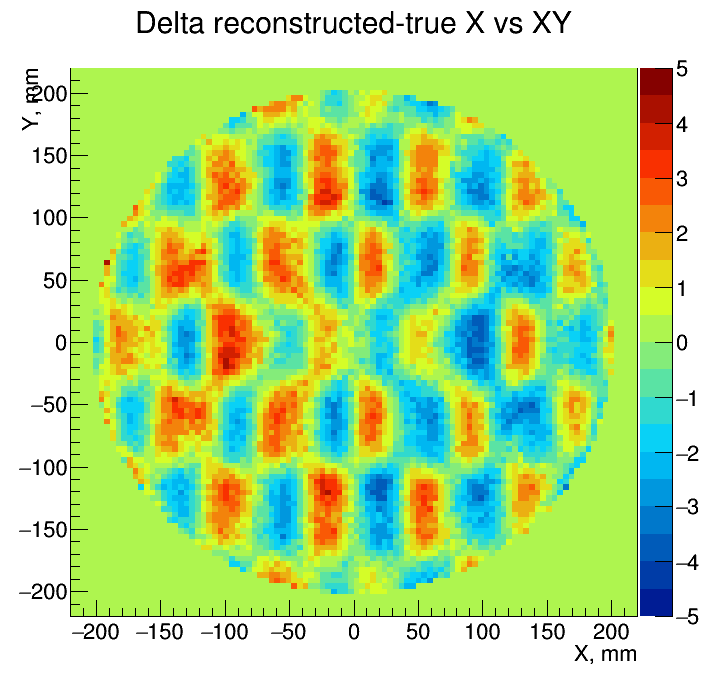} 
\includegraphics[width=4.5cm]{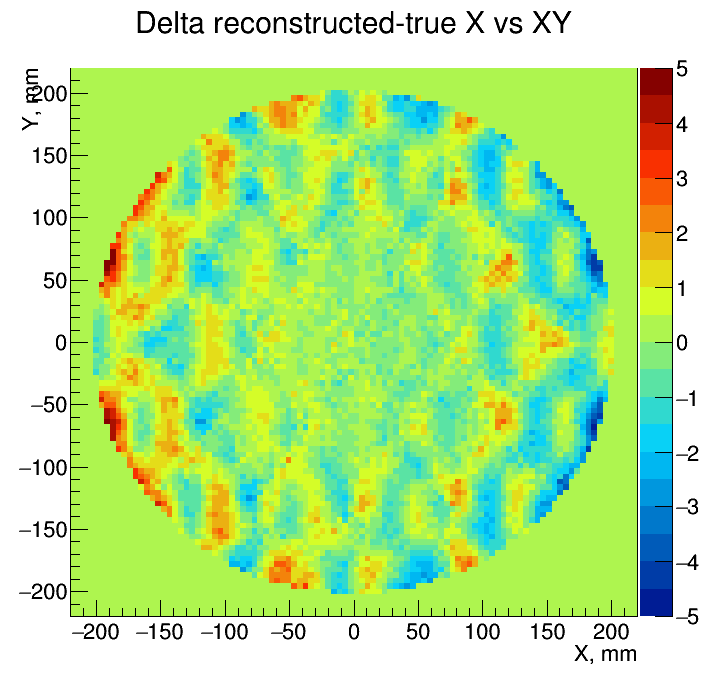}
\includegraphics[width=4.5cm]{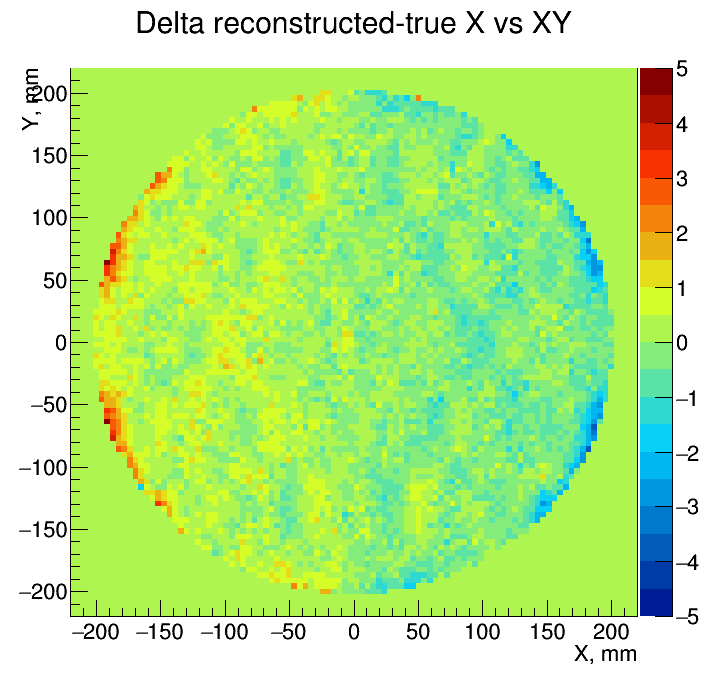}
\end{center}
\begin{center}
\includegraphics[width=4.5cm]{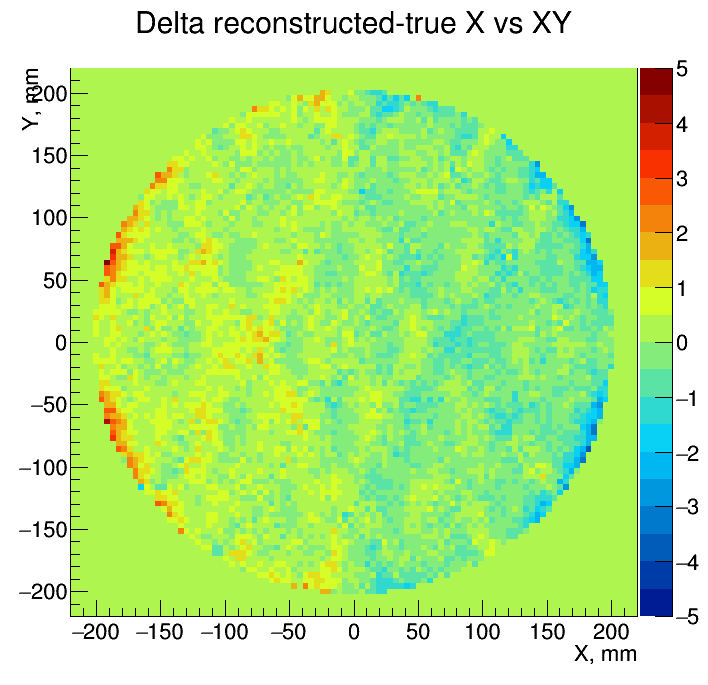} 
\includegraphics[width=4.5cm]{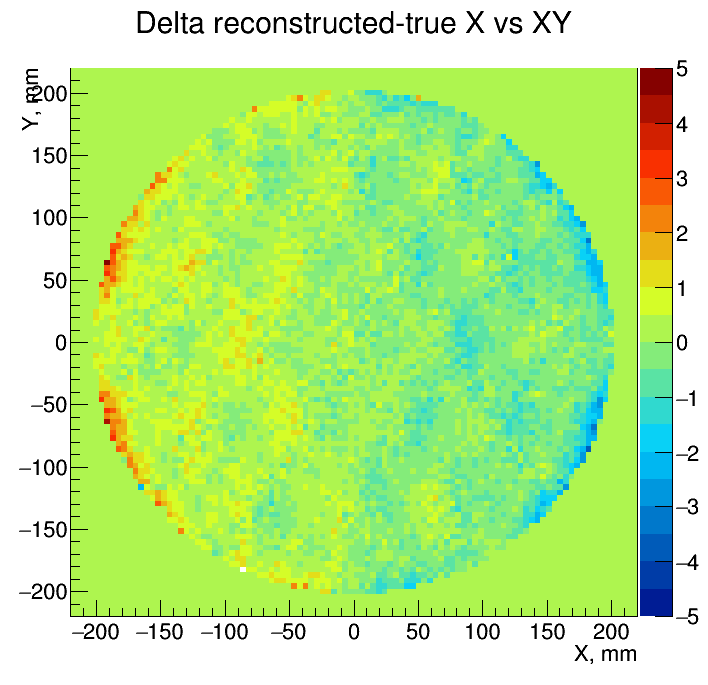}
\includegraphics[width=4.5cm]{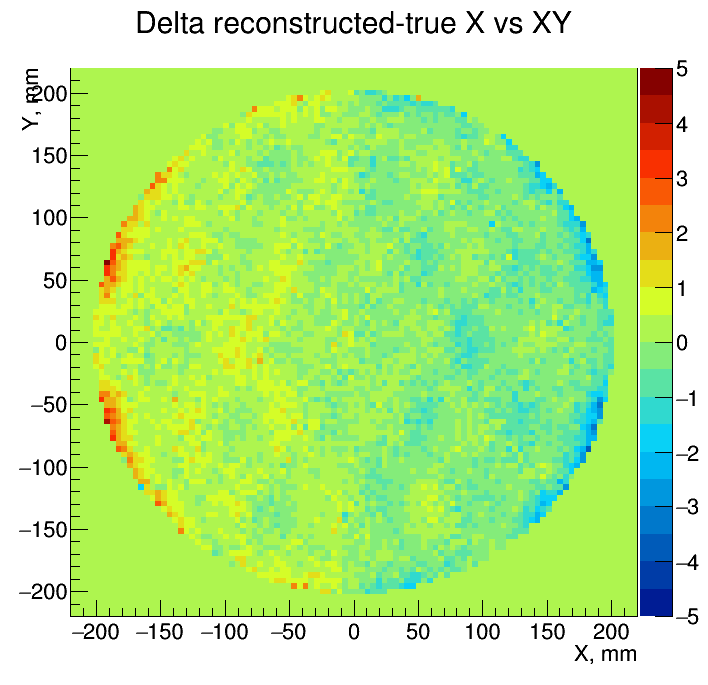}
\end{center}
\caption{\label{fig:medcam_dx}Maps of the average deviation (in mm) of the reconstructed position from the true one along the $x$ axis for a simulated uniform irradiation by monoenegetic gamma rays. The position reconstruction is performed using axial LRF parameterization with 8 (left), 12 (middle) and 16 (right) intervals per spline. Top row: no compression, bottom row: compression with $\kappa=5$, $r_0=150$mm and $\lambda=50$mm.}
\end{figure}

To quantify this result, a dataset consisting of 2$\times$10$^5$ uniform irradiation events was used to calculate the average deviation of the reconstructed position from the true one along the $x$ axis as a function of the true event position. As one can see from the results presented in Figure \ref{fig:medcam_dx}, the not compressed 8-interval parameterization produces a reconstruction bias of up to $\pm$5 mm across the whole field of view. The reason for this is that the slope of the LRFs is too steep to be adequately reproduced by an 8-interval cubic spline. As expected, the bias gradually decreases with the increase of the number of intervals used for the parameterization. On the other hand, compression results in a smoother LRF decrease (similar to the red curve in Figure \ref{fig:plot_compression}) which can be fitted by the same type of spline with a better precision. Indeed, when compression is used, the absolute bias for the whole field of view (with the exception of the narrow border region) does not exceed 2~mm for the 8-interval LRF and 1.5~mm for the 16-interval ones.  As the intrinsic spatial resolution for this camera is approximately 4~mm FWHM at 140 keV, this level of distortion is difficult to notice by eye, as confirmed by the mask images in Figure \ref{fig:masks}. 

\section{Conclusions}

A framework was developed for parameterization of the light response functions of photosensors in a scintillation camera. It is based on the approximation of the measured or simulated photosensor response using weighted sums of uniform cubic B-splines or their tensor products. The LRFs parameterized in this way are smooth, computationally inexpensive to evaluate and typically require much less storage than non-parametric alternatives. The parameters are found by the linear least squares fitting in unsupervised mode.

Both the axial symmetry of the photosensor response and the symmetry of the photosensor array can be exploited in this framework to improve stability of the parameterization and to reduce the storage requirements. An optional non-linear coordinate transform permits to further reduce the required storage space while maintaining a faithful parameterization. It can also be used as an additional LRF regularization tool allowing to avoid overfitting.

The framework is implemented in an open source software package ANTS2, designed for simulation and experimental data processing for Anger camera type detectors. Examples of the framework application are demonstrated for two broadly used types of scintillation camera geometries.

\section{Acknowledgment}

This work was carried out with financial support from the Funda\c{c}\~{a}o para a Ci\^{e}ncia e Tecnologia (FCT) through the project-grant PTDC/BBB-BMD/2395/2012 (co-financed with FEDER/COMPETE) and grant IF/00378/2013/CP1172/CT0001 as well as from Quadro de Refer\^{e}ncia Estrat\'{e}gica Nacional (QREN) in the framework of the project Rad4Life. 

\bibliographystyle{JHEP}
\providecommand{\href}[2]{#2}\begingroup\raggedright\endgroup

\end{document}